\documentclass[11pt,a4paper]{article}
\usepackage[margin=25mm,headheight=14pt,headsep=12pt]{geometry}
\usepackage{fontspec}
\usepackage{xeCJK}
\setCJKsansfont{FandolHei-Regular.otf}[BoldFont=FandolHei-Bold.otf]
\usepackage{fancyhdr}
\fancypagestyle{plain}{%
  \fancyhf{}%
  \fancyhead[L]{\normalfont\fontsize{10}{12}\selectfont 中国语音学报\quad 第24辑}%
  \fancyfoot[C]{\thepage}%
}
\usepackage{amsmath,amssymb}
\usepackage{graphicx,booktabs,array}
\usepackage[font=small,labelfont=bf]{caption}
\usepackage[section]{placeins}
\usepackage{enumitem}
\usepackage[hidelinks,unicode]{hyperref}
\hypersetup{pdftitle={Speech Emotion Recognition Across Languages: Prosodic Patterns and Self-Supervised Learning},pdfauthor={Tianqi Geng, Zhichen Han, Hui Feng, Jiahong Yuan, Korin Richmond, Yuanchao Li}}
\title{\bfseries Speech Emotion Recognition Across Languages:\\Prosodic Patterns and Self-Supervised Learning}
\author{GENG Tianqi\quad HAN Zhichen\quad FENG Hui\\YUAN Jiahong\quad RICHMOND Korin\quad LI Yuanchao\raisebox{.6ex}{\fontsize{7}{7}\selectfont *}}
\date{}
\begin{document}
\begin{center}
{\fontsize{17}{21}\selectfont\bfseries
Speech Emotion Recognition Across Languages:\\
Prosodic Patterns and Self-Supervised Learning\par}
\vspace{3pt}
{\fontsize{11}{14}\selectfont
GENG Tianqi\quad HAN Zhichen\quad FENG Hui\\
YUAN Jiahong\quad RICHMOND Korin\quad LI Yuanchao\raisebox{.6ex}{\fontsize{7}{7}\selectfont *}\par}
\end{center}
\begingroup
\renewcommand{\thefootnote}{*}
\footnotetext[0]{Corresponding author (通讯作者): \href{mailto:yl809@cam.ac.uk}{yl809@cam.ac.uk}.}
\endgroup
\vspace{0pt}
\begingroup
\fontsize{10}{12.5}\selectfont
\setlength{\parskip}{4pt}
\noindent\textbf{Abstract}\quad This study investigates the prosodic patterns of emotional speech across Chinese (CN), German (DE), and English (EN), with a focus on acoustic characteristics. Using ANOVA and mixed-effects modeling, we reveal cross-linguistic and emotion-specific prosodic patterns. A comparative analysis of Speech Emotion Recognition (SER) performance demonstrates that humans outperform Self-Supervised Learning (SSL) models in cross-lingual tasks, while SSL models show significant improvement in cross-lingual SER through an efficient fine-tuning process, achieving performance competitive with native human speakers. The study also analyzes the performance of Speech Emotion Diarization (SED) between models and humans, highlighting the effectiveness of SSL models in transfer learning scenarios. These findings emphasize the interplay between universal prosodic features (e.g., intensity, pitch) and language-specific factors. Moreover, the SSL model's outstanding performance indicates the potential value of these acoustic features in emotion processing, implying possible advancements in cross-lingual emotion recognition through refined training techniques.

\noindent\textbf{Keywords:} Prosodic Pattern; Speech Emotion; Self-Supervised Models; Speech Emotion Recognition; Speech Emotion Diarization

\vspace{2pt}
\begin{center}
{\fontsize{12}{16}\selectfont\bfseries 跨语言的语音情感识别：韵律模式与自监督学习\par}
\vspace{5pt}
{\fontsize{10}{14}\selectfont 耿天琪\quad 韩智宸\quad 冯卉\quad 袁家宏\quad Korin Richmond\quad 李远超\raisebox{.6ex}{\fontsize{7}{7}\selectfont *}\par}
\end{center}
\fontsize{10}{13.5}\selectfont
\setlength{\parindent}{2em}
\noindent\textbf{摘要：}本研究旨在探讨汉语、英语和德语中语音情感表达的韵律模式及其间的跨语言差异。通过使用单因素方差分析和混合效应模型，本研究系统分析了情感对相关声学参数的影响，并在控制性别和语言差异的情况下，揭示跨语言情况下情感的特定韵律表达。研究结果显示，不同语言在情感韵律表达上存在一定的共性，但同时也展现出语言类型和文化背景等不同所导致的差异。具体而言，本文所使用的汉语和德语语料均在不同情感的表达中表现出对强度相关特征的普遍依赖，而英语语料则不然。基频方面，研究发现开心情感的基频普遍最高，愤怒次之，而悲伤情绪的基频最低。然而，混合效应模型的结果显示，情感对韵律表达的影响在不同文化和语言环境下存在差异，表明语言和文化背景在情感韵律的解读中扮演了重要角色。

在此基础上，本研究还对跨语言的语音情感识别的性能进行了比较分析，重点探讨了人类被试和自监督学习（Self-Supervised Learning，SSL）模型在跨语言情感识别任务中的表现差异。实验结果表明，人类在跨语言情感识别任务中通常优于 SSL 模型。通过参数高效微调（Parameter-Efficient Fine-Tuning）策略，SSL 模型在语音情感识别任务中的表现有了显著提升。在跨语言任务中，模型的识别能力逐渐接近母语使用者的水平。这一发现证明了 SSL 模型在跨语言的语音情感识别中的潜力。

此外，研究还进行了语音情感分段识别（Speech Emotion Diarization，SED）任务，重点分析了语音情感片段的精细化识别。研究表明，SSL 模型在识别情感片段时能够通过一种语言上的训练，掌握其特定的情感模式，在该语言以及多语言的背景下，模型对情感特征的敏感度和识别准确度有了明显提高。对人类被试所切分出的最具代表性的语音情感片段作进一步的韵律特征分析，结果表明强度作为情感感知的普遍线索在不同语言中都具有重要性。同时，研究还发现，基频相关的特征在三种语言的情绪感知中展现出跨语言的共性。

本研究强调了人类感知与 SSL 模型在语音情感识别任务中的异同。尽管 SSL 模型在情感识别中取得了一定的进展，但人类感知往往涉及更多的语境和语义信息，这些是当前 SSL 模型尚未完全模拟的。未来的研究可以进一步探讨如何将更多的语境和语义信息整合到 SSL 模型中，以提高其情感识别的精度。同时，也需要更多样化的数据集以及对发音人、文化背景和语言类型等变量的控制，从而进一步优化和细化本研究的结论。

\noindent\textbf{关键词：}韵律模式；语音情感；自监督模型；语音情感识别；语音情感分段识别
\par
\endgroup
\clearpage

\section{Introduction}

\subsection{Prosody and emotional speech}

Emotions are generally classified as different types using discrete categories, such as the ``Big Six'' (Ekman et al., 1969), or continuous dimensions, such as the Circumplex Model (Russell, 1980), the Valence-Arousal-Dominance framework (Mehrabian, 1996), and the FEELTRACE scale (Cowie et al., 2000). When we speak, we use speech prosody and nonlinguistic vocalizations, along with concrete words, to convey not only linguistic information but also nonlinguistic information, such as emotion.

Speech emotions have attracted significant attention from researchers, with studies examining both their production and perception (Hancil, 2009; Mac et al., 2012). In terms of production, speech prosody is recognized as playing a central role in conveying emotions. Prosodic features, such as fundamental frequency, speech rate, and intensity, are closely associated with emotional states (Banse \& Scherer, 1996; Williams \& Stevens, 1972). Previous studies have also shown that the perception of emotion in Mandarin is multi-cued, while different emotions may rely on different perceptual cues. In particular, mean duration and mean intensity have been found to play a substantial role in the identification of sadness and anger (Zhang et al., 2020). Moreover, prosodic communication of emotion can operate independently of verbal content (Kitayama \& Ishii, 2002) and relies on similar cues across different cultures (Bolinger, 1978).

On the perception side, experimental studies reveal that a wide variety of positive and negative emotions can be recognized from vocal expressions (Cowen et al., 2019). Paralinguistic features enable listeners to recognize discrete emotions across spoken languages (Pell, Monetta, et al., 2009; Scherer et al., 2001). Interestingly, although cross-linguistic differences in the perception of emotional speech have been observed (Elfenbein \& Ambady, 2003; Erickson, 2010; Pell, Monetta, et al., 2009), listeners can accurately identify emotions not only from their own language but also from other languages with recognition accuracies surpassing chance performance levels (Pell, Paulmann, et al., 2009; Thompson \& Balkwill, 2006), demonstrating the universality in the recognition of speech emotions (Jeon et al., 2013).

\subsection{Speech emotion recognition}

Powerful pre-trained speech models, such as Wav2vec 2.0 (W2V2) (Baevski et al., 2020) and WavLM (Chen et al., 2022), as well as their multilingual variants, have been developed as Self-Supervised Learning (SSL) continues to advance.

These models have demonstrated remarkable performance in a range of downstream speech tasks, including speech emotion recognition (SER) (Li et al., 2022). To further improve the applicability of SSL models across different languages and datasets for SER, Parameter-Efficient Fine-Tuning (PEFT) has been utilized to enhance their performance while reducing the necessity for fine-tuning an entire model (Feng \& Narayanan, 2023; Lashkarashvili et al., 2024).

However, linguistic and cultural disparities remain a significant obstacle to cross-lingual SER (Shoukat et al., 2023). Improving cross-lingual SER performance often requires large amounts of training data in the target language, which can be challenging for languages with limited emotional speech datasets (Ahn et al., 2021; Neumann, 2018).

Previous research has primarily relied on target language data to refine SER models; however, when transitioning from monolingual to cross-lingual settings, a notable decline in performance has been observed (Latif et al., 2018; Neumann, 2018). Additionally, adversarial neural networks have been explored for cross-lingual adaptation in uncontrolled environments (Cai et al., 2021; Latif et al., 2019). More recently, Upadhyay et al. (2024) established a layer-anchoring approach to facilitate emotion transfer, considering the hierarchical structure and task-specific nature of speech models.

\subsection{Research questions}

While Self-Supervised Learning models have demonstrated potential in cross-lingual speech emotion recognition (Singh \& Gupta, 2023), two critical gaps remain: a) insufficient understanding of how acoustic-prosodic features mediate cross-linguistic emotion communication, and b) the lack of systematic comparison between machine and human perceptual patterns. By examining how different emotions are interpreted through acoustic features such as pitch, duration, and intensity, this study aims to identify both universal and language-specific patterns in emotional prosody. Furthermore, this study aims to compare SSL models and human performance by investigating fine-grained emotion perception (i.e., segment-level SER) and fine-tuning techniques in addition to traditional utterance-level SER.

To this end, we raise the following research questions:

\begin{enumerate}[label=\arabic*)]

\item What are the prosodic features of different emotions in Mandarin Chinese, English, and German?

\item How do human listeners leverage these prosodic features to recognize emotions, particularly when processing non-native speech?

\item Can SSL-based models achieve SER performance comparable to that of humans?

\item How can SSL models for SER be more effectively fine-tuned in cross-lingual scenarios?

\item Can SSL-based models identify emotionally salient segments in a manner that is similar to human behavior?

\end{enumerate}

\section{Methodology}

Jeon et al. (2013) discovered that while cross-lingual problems had less impact on people, Support Vector Machine models performed better in monolingual contexts. Further research by Werner and Petrenko (2019) revealed that human cross-lingual skills in SER are often strong. Despite these advancements, there is still a lack of human-model comparison research, leaving this field understudied.

\subsection{Datasets and models}

Four public emotion corpora were used:

\begin{itemize}

\item ESD: Ten native Mandarin Chinese (CN) speakers, including five men and five women, produced utterances in five different emotion categories in a Mandarin Chinese emotion corpus (Zhou et al., 2022).

\item PAVOQUE: A German (DE) emotion corpus (Steiner et al., 2013) with five emotion categories produced by a professional male actor. Inside the dataset, speech with neutral emotion accounts for over a half.

\item IEMOCAP: An English (EN) emotion corpus (Upadhyay et al., 2024) contains nine emotion categories from both scripted and improvised emotional conversations involving five male and female speaker pairs.

\item ZED: a corpus of English emotion data created especially for the SED task (Wang et al., 2023), which includes human annotations of speech data at the utterance and sub-utterance (segment) levels. The corpus contains single-channel 16kHZ WAV files converted from YouTube videos, covering both male and female speech in comedy shows, victim interviews, sports commentary, and beyond.

\end{itemize}

We employed a WavLM-large model trained on EN emotional speech\footnote{\url{https://huggingface.co/speechbrain/emotion-diarization-wavlm-large}} in addition to three W2V2-base models pre-trained on CN\footnote{\url{https://huggingface.co/TencentGameMate/chinese-wav2vec2-base}}, DE,\footnote{\url{https://huggingface.co/facebook/wav2vec2-base-de-voxpopuli-v2}} and EN\footnote{\url{https://huggingface.co/facebook/wav2vec2-base-960h}} (Wang et al., 2023). The following activities were carried out for certain objectives utilizing various models and datasets.

\subsection{Subjects}

Two groups of subjects were recruited for the test. The first group consists of six native Mandarin Chinese speakers (two males and four females) who have never learned German. The second group consists of six native German speakers (one male and five females) who have never learned Chinese. Both groups are highly proficient in English, and none have impairments in hearing or emotional comprehension.

\subsection{Model training}

\subsubsection{Layer-wise analysis of SSL models}

To explore monolingual and cross-lingual SER performances, three datasets were used: ESD (CN), PAVOQUE (DE), and IEMOCAP (EN). The emotions analyzed include angry, happy, neutral, and sad. For the models, we employed W2V2-CN, -DE, and -EN, representing SSL models pre-trained on Chinese, German, and English speech data, respectively.

SSL models encode speech information across different layers, and prior research has shown that representations from the middle layers often yield superior performance in SER tasks (Li et al., 2023). To determine the optimal layer for both monolingual and cross-lingual SER, we performed a layer-wise analysis. In this approach, SSL models were used as feature extractors with all parameters frozen, and Unweighted Accuracy (UA)\footnote{Also known as the overall accuracy, which calculates the fraction of instances predicted correctly, without considering class distribution.} is used as the evaluation metric.

The analysis includes three experimental settings. In the monolingual setting, models were fine-tuned and tested on data in the same language as their pre-training language. For example, W2V2-CN was fine-tuned and tested using Chinese data (ESD). In the cross-lingual setting, models were fine-tuned using data in their pre-training languages but tested in a different language. For instance, W2V2-CN was fine-tuned on Chinese data (ESD) and tested on German data (PAVOQUE) or English data (IEMOCAP). Lastly, in the transfer learning setting, models were fine-tuned and tested on data in a language different from their pre-training language. For example, W2V2-CN was fine-tuned and tested on German (PAVOQUE) or English (IEMOCAP) data.

\subsubsection{PEFT of SSL models for cross-lingual SER}

Following the layer-wise analysis, the most effective layers were further optimized using several PEFT methods. Specifically, we employed the Low-Rank Adapter (LoRA) (Hu et al., 2021), Bottleneck Adapter (BA) (Houlsby et al., 2019), and Weighted Gating (WG) techniques. In addition, a two-stage fine-tuning process (Lashkarashvili et al., 2024) was applied, in which the model was first fine-tuned on the source language and then fine-tuned on the target language after the initial stage achieved convergence.

\subsection{Experiment procedure}

\subsubsection{Prosodic analysis}

To examine the prosodic pattern of each specific emotion, all the audio materials were automatically segmented by xSegmenter (Xiong, 2019) and manually corrected in Praat. The average $f_0$ (in Hz), together with $f_0$ minimum, $f_0$ maximum, duration, and mean intensity, was extracted through a Praat script. All $f_0$-related measures were normalized using a semitone transformation (Shi et al., 2009), whereas non-$f_0$ parameters were kept in their original scales. Speech rate was computed as the number of words divided by sentence duration (ms) while Int\_SD was calculated as the standard deviation of mean intensity.

To examine the differences in terms of acoustic features between selected and unselected segments from SED tasks, all acoustic parameters were analyzed in their unnormalized forms. Given that this analysis focuses on contrasts within groups, normalization was unnecessary and would not contribute to the validity of the comparison.

One-way ANOVA was conducted to examine the effect of emotion on the acoustic characteristics. To further control for individual differences across languages and genders, we established a linear mixed-effects model as follows:

\begin{equation}\mathrm{Parameter}\sim\mathrm{emotion}+(1\mid\mathrm{gender})+(1\mid\mathrm{language}).\label{eq:mixed}\end{equation}

In this model, emotion was treated as a fixed effect, while language and gender were considered random effects to control for their potential impact on the parameter. The emotion category ``neutral'' was set as the reference, allowing assessment of the overall impact of emotion on the parameter while accounting for variability due to language and gender, ensuring the robustness of the results.

\subsubsection{SER task}

This task investigated speech emotion recognition using three datasets, ESD, PAVOQUE, and IEMOCAP, along with all the selected models. Four emotions-angry, happy, neutral, and sad were considered.

As for model training, 400 utterances per emotion were selected, amounting to 1600 utterances per dataset. Validation and test sets each consisted of 200 randomly selected utterances. Five-fold cross-validation was applied for model training to enhance reliability. An equal amount of data was used for the three languages to minimize the impact of variations in training data size while each dataset was balanced across the four emotion categories, ensuring an equal distribution of samples per emotion.

In layer-wise analysis, a classification head that projects from a dimension of 768 to 4 was utilized. The training parameters included a learning rate of 1e-4, an epsilon value of 1e-8, and a weight decay of 1e-5. The model was trained for 100 epochs with a batch size of 32. The loss criterion used was cross-entropy. Training would be stopped if the validation loss did not improve for 10 consecutive epochs. In the PEFT configuration, the same classification head setup remained as used in the layer-wise analysis. For the LoRA module, we set the number of attention heads to 8, the scaling factor (alpha) to 16, and applied a dropout rate of 0.1. In the BA module, the reduction factor was configured to 16. The models were trained for 100 epochs with a batch size of 16. The loss and stopping criteria from the layer-wise analysis were preserved.

As for the human perception experiment, 12 sentences were chosen from each emotion category, resulting in a total of 144 utterances (12 sentences * 4 emotions * 3 datasets) presented as stimuli. By addressing the web-MUSHRA interface (Schoeffler et al., 2018), participants were tasked with listening to these speech samples and identifying the emotions they convey. Consistent with model performance evaluation, UA was used as the evaluation metric.

\subsubsection{SED task}

In this task, we utilized the ESD, PAVOQUE, and ZED datasets and models. The emotions considered are angry, happy, and sad, with neutral emotion excluded due to its lack of perceptible emotional variation for segmentation. To explore fine-grained speech emotion expression, we performed Speech Emotion Diarization (SED), where participants first listened to speech samples and labelled the emotion, as done in the SER task. Afterward, they clipped the speech and selected the segment that most clearly expresses the emotion. We adopted the Emotion Diarization Error Rate (EDER) as the evaluation metric (Wang et al., 2023). This metric calculates the error rate of diarization results, accounting for missed emotions (ME), false alarms (FA), overlaps (OL), and confusion (CF). A lower score indicates better performance.

\begin{equation}\mathrm{EDER}=\frac{\mathrm{ME}+\mathrm{FA}+\mathrm{OL}+\mathrm{CF}}{\text{Utterance Duration}}.\label{eq:eder}\end{equation}

Given the considerable effort required for segmenting speech, the model performed the SED task exclusively on English utterances from the ZED dataset, with 8 utterances randomly selected per emotion, resulting in a total of 24 utterances (8 *3 emotions). Meanwhile, human participants completed the task across all three languages, evaluating a total of 72 utterances (24*3 languages), with the additional utterances for Chinese and German sourced from the ESD and PAVOQUE datasets.

\section{Results and discussion}

\subsection{Prosodic pattern}

\subsubsection{Fundamental frequency}

Our corpora were sourced from open datasets for model training, inevitably resulting in some inconsistencies in data uniformity. Given this limitation, subsequent analyses and investigations into acoustic characteristics will primarily focus on within-language comparisons.

\paragraph{1) $f_0$ Mean}

\begin{figure}[!htbp]
\centering
\includegraphics[width=.66\linewidth]{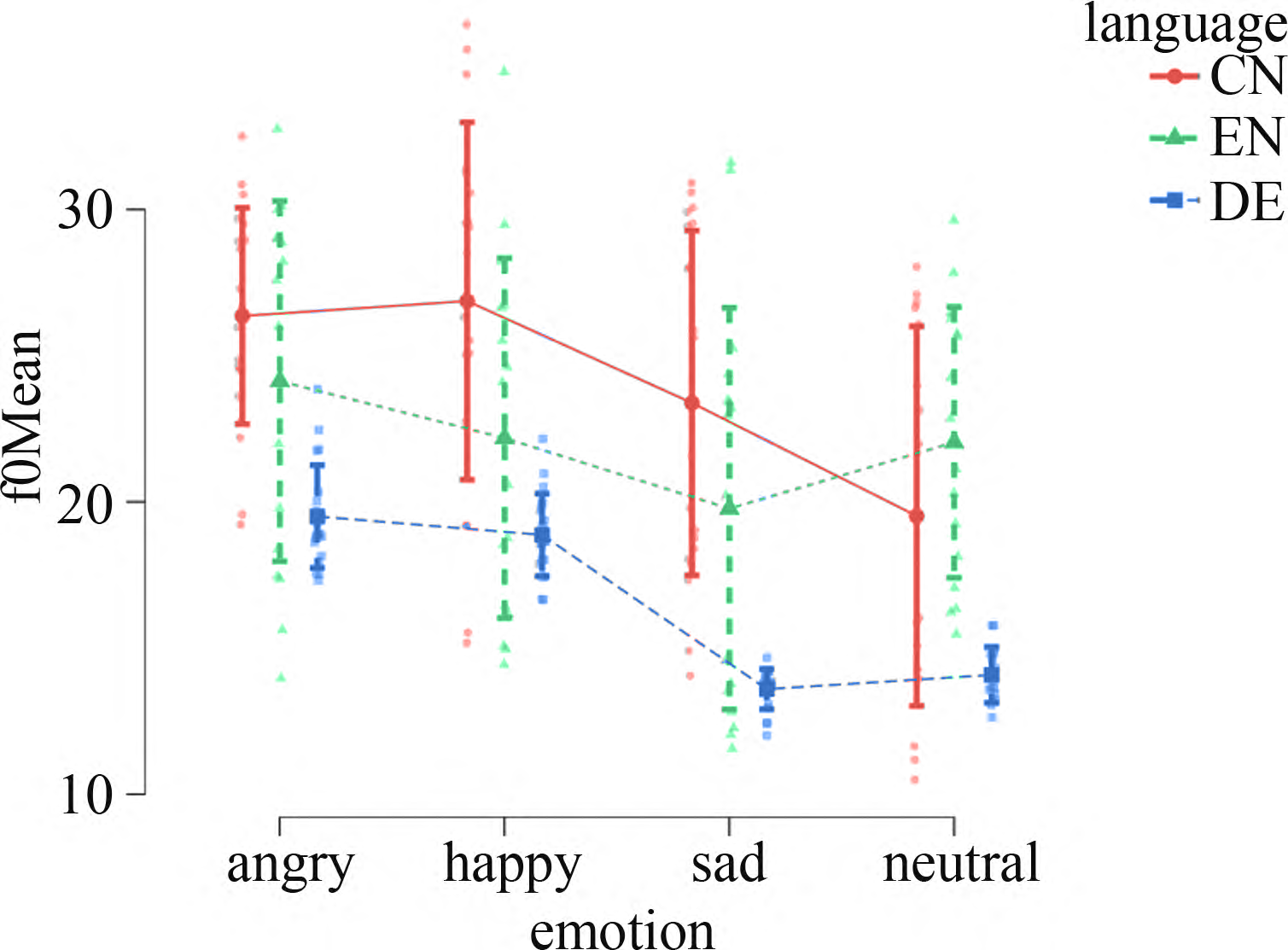}
\caption{$f_0$ mean in Chinese, German, and English by emotion.}
\label{fig:1}
\end{figure}

From Figure~\ref{fig:1}, it is clear that in the Chinese dataset, the mean fundamental frequency ($f_0$) is highest for the happy emotion and lowest for the neutral emotion. In English and German, the mean $f_0$ is highest for the angry emotion and lowest for the sad emotion. Comparatively, the overall mean $f_0$ in Chinese is greater than that in English and is even greater than that in German. The mean $f_0$ in German and English exhibits a more similar pattern of emotional influence, with the order from highest to lowest being angry, happy, neutral, and sad. The observed similarity in production patterns between English and German is likely attributable to their shared typological features. In contrast, Chinese, as a tonal language, may require speakers to exert additional efforts to encode emotional expressions into the inherent tonal structure of characters.

The ANOVA results, as shown in Table~\ref{tab:1}, indicate that emotion significantly affected the $f_0$ Mean in both Chinese and German. However, no significant effect of emotion on the $f_0$ Mean was observed in English.

The results of the mixed-effects model revealed a significant intercept (\ensuremath{\beta} = 20.14, SE = 4.91, t = 4.10, p \ensuremath{<} 0.001). Compared to the neutral baseline, angry (\ensuremath{\beta} = 4.81, SE = 0.55, t = 8.69, p \ensuremath{<} 0.001) and happy (\ensuremath{\beta} = 4.55, SE = 0.55, t = 8.22, p \ensuremath{<} 0.001) significantly increased $f_0$ Mean. However, sad did not significantly differ from neutral (\ensuremath{\beta} = 0.39, SE = 0.55, t = 0.70, p = 0.485).

Random effects analysis revealed variance attributable to language ($\tau_{00}$ = 2.53) and gender ($\tau_{00}$ = 46.16), with an ICC of 0.86, indicating that differences across languages and genders can explain 86\% of the total variance in $f_0$ Mean. The residual variance $\sigma^2$ was 7.91. Marginal $R^2$ was 0.082, while conditional $R^2$ was 0.872, reflecting the substantial contribution of random effects to the model's explanatory power.

\begin{table}[!htbp]
\centering
\caption{ANOVA Results for the Effect of Emotion on $f_0$ Mean Across Languages.}
\label{tab:1}
\small
\renewcommand{\arraystretch}{1.12}
\setlength{\tabcolsep}{10pt}
\begin{tabular}{lrrrr}
\toprule
language & sumsq & df & statistic & p.value \\
\midrule
CN & 601 & 3 & 6.37 & \ensuremath{<}0.01 \\
DE & 554 & 3 & 111 & \ensuremath{<}0.01 \\
EN & 153 & 3 & 1.41 & 0.249 \\
\bottomrule
\end{tabular}
\end{table}

To further investigate the specific patterns of emotional speech across these languages, we analyze the minimum ($f_0$ Min) and maximum ($f_0$ Max) pitch values.

\paragraph{2) $f_0$ Min}

\begin{figure}[!htbp]
\centering
\includegraphics[width=.66\linewidth]{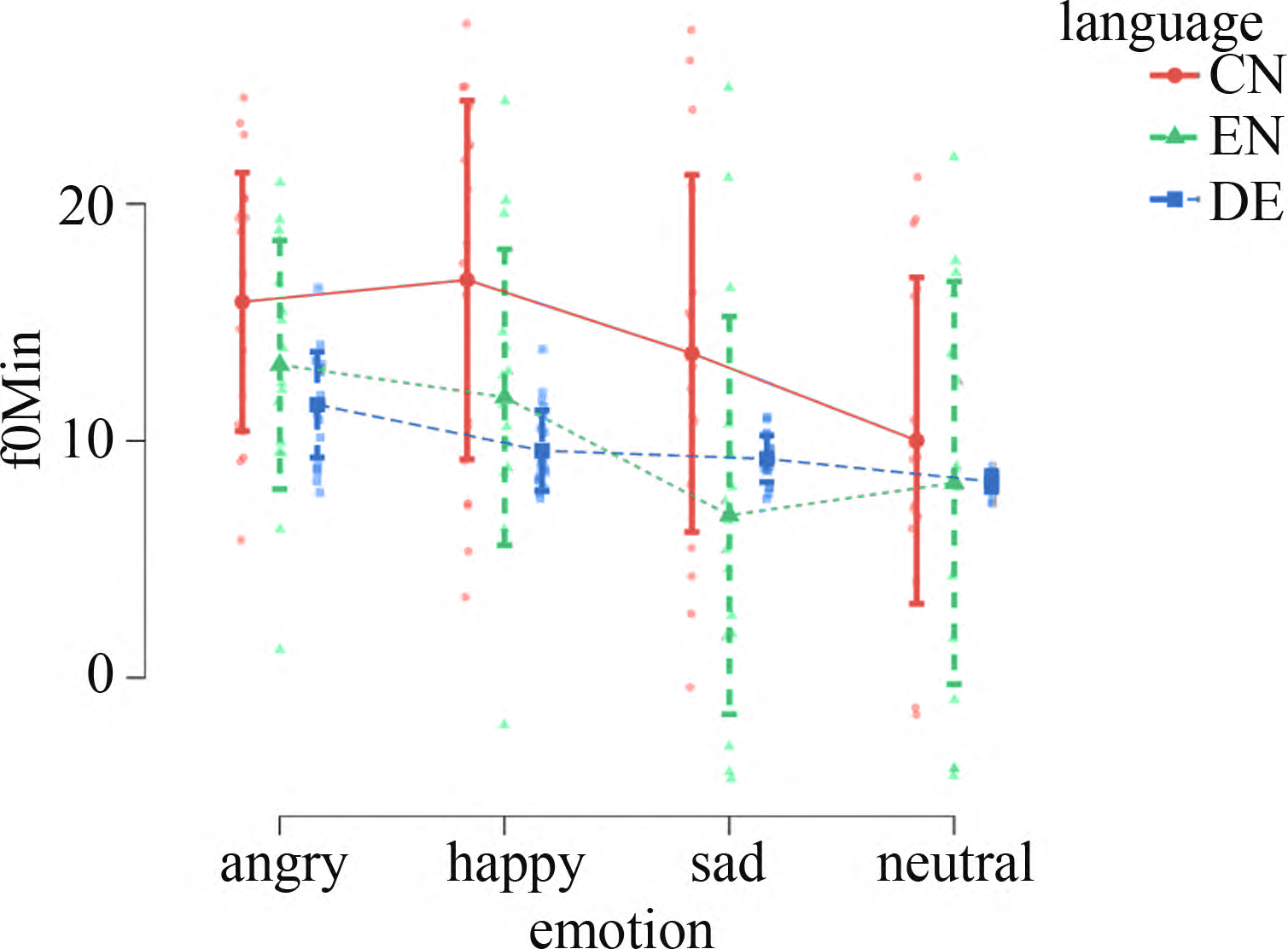}
\caption{$f_0$ Min in Chinese, German, and English by emotion.}
\label{fig:2}
\end{figure}

From Figure~\ref{fig:2}, it is evident that in the Chinese and German datasets, the $f_0$ Min is highest for the happy emotion, followed by angry, sad, and finally neutral, which shows the lowest value. In contrast, the English dataset exhibits a descending order of $f_0$ Min as follows: angry, happy, neutral, and sad. Similarly, the German dataset ranks the $f_0$ Min in the order of angry, happy, sad, and neutral. Overall, the Chinese dataset demonstrates consistently higher minimum pitch values compared to both the English and German datasets.

A one-way ANOVA was performed, as shown in Table~\ref{tab:2}, to investigate the influence of emotional expression on $f_0$ Min across three distinct languages. The results revealed a significant effect of emotion on $f_0$ Min in Chinese and German. However, no significant effect was observed in English, although the result approached significance.

\begin{table}[!htbp]
\centering
\caption{ANOVA Results for the Effect of Emotion on $f_0$ Min Across Languages.}
\label{tab:2}
\small
\renewcommand{\arraystretch}{1.12}
\setlength{\tabcolsep}{10pt}
\begin{tabular}{lrrr}
\toprule
language & sumsq & statistic & p.value \\
\midrule
CN & 475 & 3.31 & \ensuremath{<}0.05 \\
DE & 103 & 14.5 & \ensuremath{<}0.001 \\
EN & 428 & 2.72 & 0.0521 \\
\bottomrule
\end{tabular}
\end{table}

The results of the mixed-effects model revealed a significant intercept (\ensuremath{\beta} = 10.09, SE = 4.03, t = 2.50, p = 0.013). Compared to the neutral baseline, angry (\ensuremath{\beta} = 4.64, SE = 0.96, t = 4.83, p \ensuremath{<} 0.001) and happy (\ensuremath{\beta} = 4.15, SE = 0.96, t = 4.32, p \ensuremath{<} 0.001) significantly increased $f_0$ Min. However, sad does not significantly differ from neutral (\ensuremath{\beta} = 1.23, SE = 0.96, t = 1.28, p = 0.203). Random effects analysis showed variability attributable to both language ($\tau_{00}$ = 4.82) and gender ($\tau_{00}$ = 28.30), with an ICC value of 0.58, indicating that 58\% of the variance in $f_0$ Min can be attributed to differences across languages and genders.

\paragraph{3) $f_0$ Max}

\begin{figure}[!htbp]
\centering
\includegraphics[width=.66\linewidth]{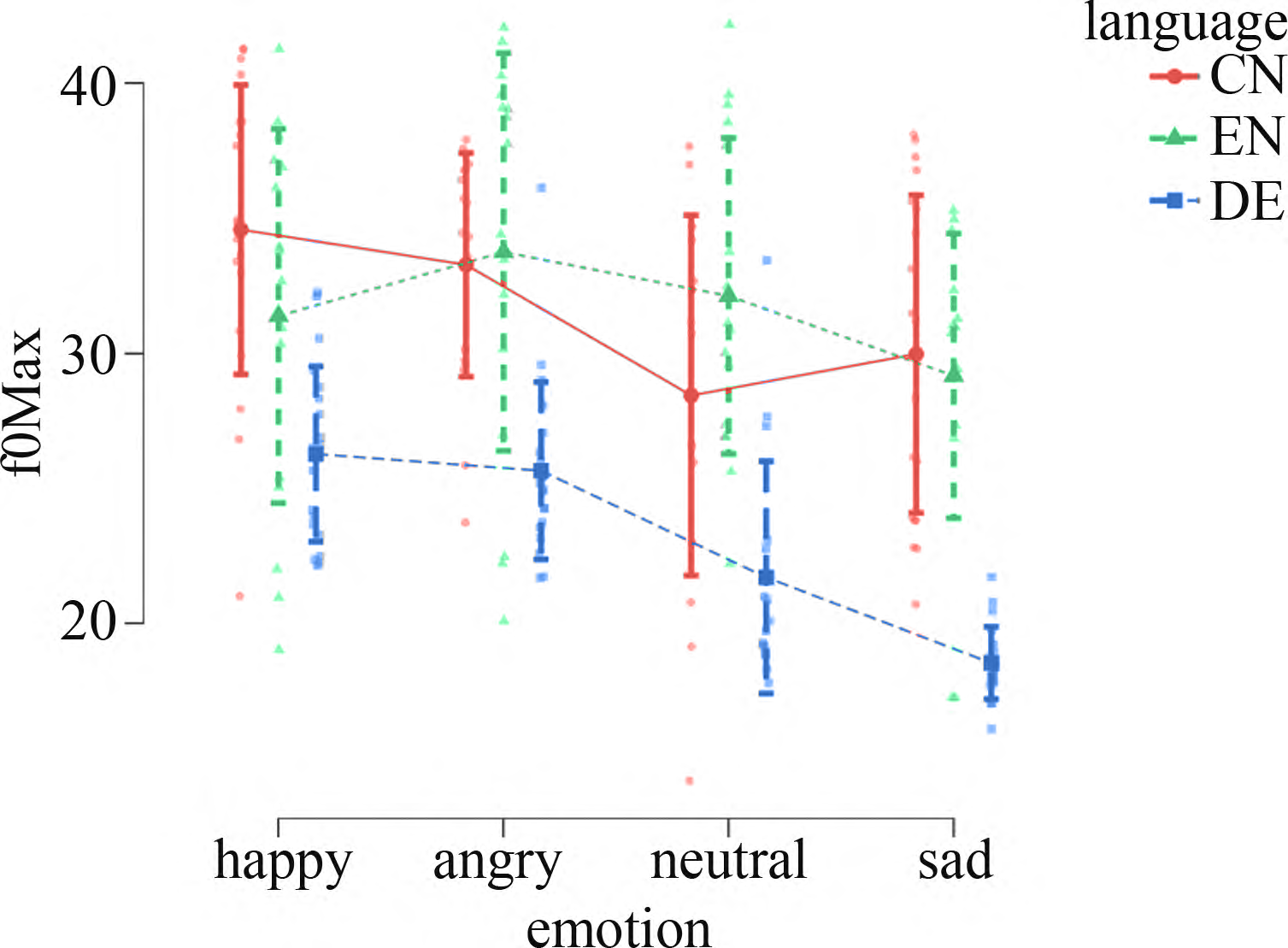}
\caption{$f_0$ Max in Chinese, German, and English by emotion.}
\label{fig:3}
\end{figure}

The analysis of $f_0$ Max across languages, as demonstrated in Figure~\ref{fig:3}, reveals distinct cross-linguistic patterns in emotional expression. In Chinese, $f_0$ Max is highest for the happy emotion, followed by angry, sad, and neutral. In English, angry shows the highest $f_0$ Max, followed by neutral, happy, and sad. In contrast, German displays a different order, with happy ranking highest, followed by angry, neutral, and sad.

Table~\ref{tab:3} summarizes the results of the one-way ANOVA examining the effect of emotion on $f_0$ Max. Significant effects of emotion on $f_0$ Max were observed in both Chinese and German, indicating that emotional categories influence $f_0$ Max in these languages. However, no significant effect was found in English, suggesting that emotional variation does not substantially impact $f_0$ Max in this language. These findings reflect stronger effects in tonal and stress languages, such as Chinese and German, compared to English.

\begin{table}[!htbp]
\centering
\caption{ANOVA Results for the Effect of Emotion on $f_0$ Max Across Languages.}
\label{tab:3}
\small
\renewcommand{\arraystretch}{1.12}
\setlength{\tabcolsep}{10pt}
\begin{tabular}{lrrr}
\toprule
language & sumsq & statistic & p.value \\
\midrule
CN & 445 & 4.87 & \ensuremath{<}0.05 \\
DE & 779 & 26.1 & \ensuremath{<}0.001 \\
EN & 174 & 1.41 & 0.248 \\
\bottomrule
\end{tabular}
\end{table}

The results of the mixed-effects model for $f_0$ Max revealed a significant intercept (\ensuremath{\beta} = 28.69, SE = 4.19, t = 6.85, p \ensuremath{<} 0.001). Compared to the neutral baseline, angry (\ensuremath{\beta} = 3.46, SE = 0.83, t = 4.17, p \ensuremath{<} 0.001) and happy (\ensuremath{\beta} = 3.74, SE = 0.83, t = 4.51, p \ensuremath{<} 0.001) significantly increased $f_0$ Max. However, sad did not significantly differ from neutral (\ensuremath{\beta} = -1.57, SE = 0.83, t = -1.90, p = 0.059), showing a trend toward lower $f_0$ Max, although it did not reach statistical significance.

Random effects analysis indicated variance attributable to both language ($\tau_{00}$ = 7.81) and gender ($\tau_{00}$ = 29.10), with an ICC of 0.68, suggesting that differences across languages and genders can account for 68\% of the total variance in $f_0$ Max. The residual variance was 17.73. Marginal $R^2$ was 0.088, while conditional $R^2$ was 0.704, reflecting that random effects accounted for a substantial portion of the model's explanatory power.

\subsubsection{Intensity}

\begin{figure}[!htbp]
\centering
\includegraphics[width=.66\linewidth]{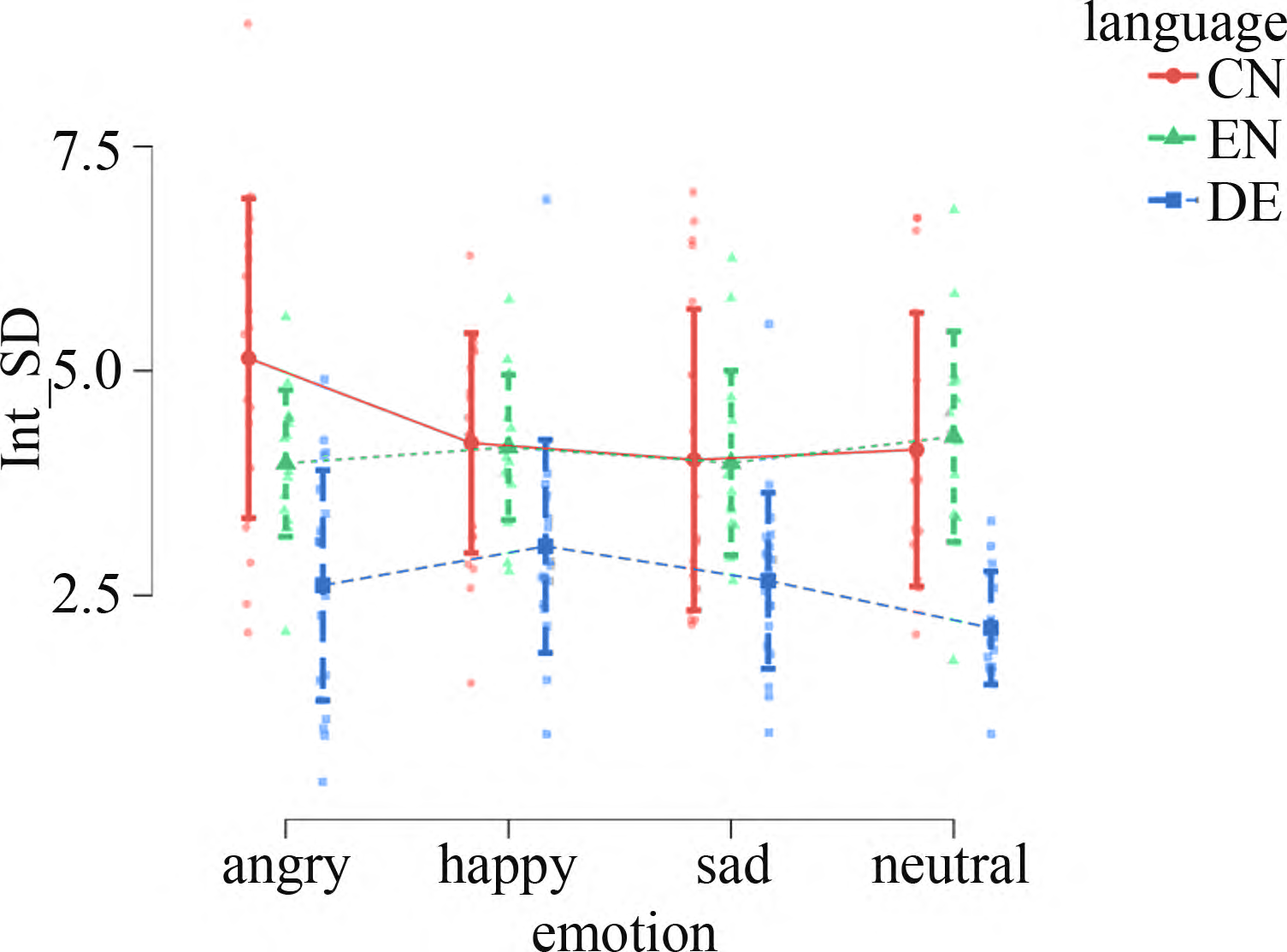}
\caption{Int\_SD in Chinese, German, and English under by emotion.}
\label{fig:4}
\end{figure}

The Int\_SD, calculated as the standard deviation of the average intensity values of syllables within each sentence, reflects the degree of variation in the intensity of each sentence. Higher Int\_SD values indicate greater intensity fluctuations, while lower values suggest more stable intensity. As Fig.~\ref{fig:4} shows, in the Chinese dataset, the widest intensity fluctuation range occurred under the angry emotion, followed by happy, neutral, and the narrowest range under sad. In the English dataset, the neutral emotion exhibited the widest range, followed by happy, with angry and sad showing similar fluctuation ranges. In the German dataset, the widest fluctuation range was observed under happy, the narrowest under neutral, while angry and sad exhibited similar intensity fluctuation patterns.

Since we are comparing the standard deviation of intensity, it is reasonable to treat this measure as comparable across the three languages. Overall, German exhibited the narrowest degree of intensity fluctuation, while Chinese and English demonstrated similar patterns across happy, sad, and neutral emotions. It can be concluded that, despite their distinct typologies, Chinese and English speakers tend to use similar intensity strategies when marking emotions, with the notable exception of anger. For angry, Chinese speakers exhibit greater variation in intensity compared to English speakers. In contrast, the significant difference observed between German and the other two languages may be attributed to several factors. First, the German dataset was primarily recorded by a single male speaker, who may possess a more introverted personality, or the dataset may inherently reflect a more stable representation of intensity. Additionally, the cultural background of German speakers could also influence these results. These possibilities warrant further exploration in future research, potentially involving more diverse stimuli and speakers.

\begin{table}[!htbp]
\centering
\caption{ANOVA Results for the Effect of Emotion on Int\_SD Across Languages.}
\label{tab:4}
\small
\renewcommand{\arraystretch}{1.12}
\setlength{\tabcolsep}{10pt}
\begin{tabular}{lrrr}
\toprule
language & sumsq & statistic & p.value \\
\midrule
CN & 16 & 2.17 & 0.0986 \\
DE & 7.37 & 2.15 & 0.101 \\
EN & 1 & 0.357 & 0.785 \\
\bottomrule
\end{tabular}
\end{table}

However, the results of the ANOVA test shown in Table~\ref{tab:4} indicate that the influence of emotion on Int\_SD was not statistically significant for any of the three languages. In the Chinese dataset, emotion exhibited a marginal effect, suggesting a trend toward significance. Similarly, for the German dataset, the effect of emotion approached significance. However, in the English dataset, no significant effect of emotion was observed. These results suggest that the variability in the fluctuation of intensity was not strongly modulated by emotion across the analyzed languages. Meanwhile, since ANOVA revealed no significant effect, the mixed-effects model did not proceed, as the primary hypothesis regarding the influence of emotion was not supported.

\subsubsection{spRate}

Considering the various types of languages, the calculation of speech rate (spRate) in this study differs by language: for Chinese, it is derived by dividing the total duration of the sentence by the number of syllables, while for English and German, it is calculated by dividing the total duration by the number of words. A larger spRate indicates slower speech, while a smaller value signifies faster speech.

\begin{figure}[!htbp]
\centering
\includegraphics[width=.66\linewidth]{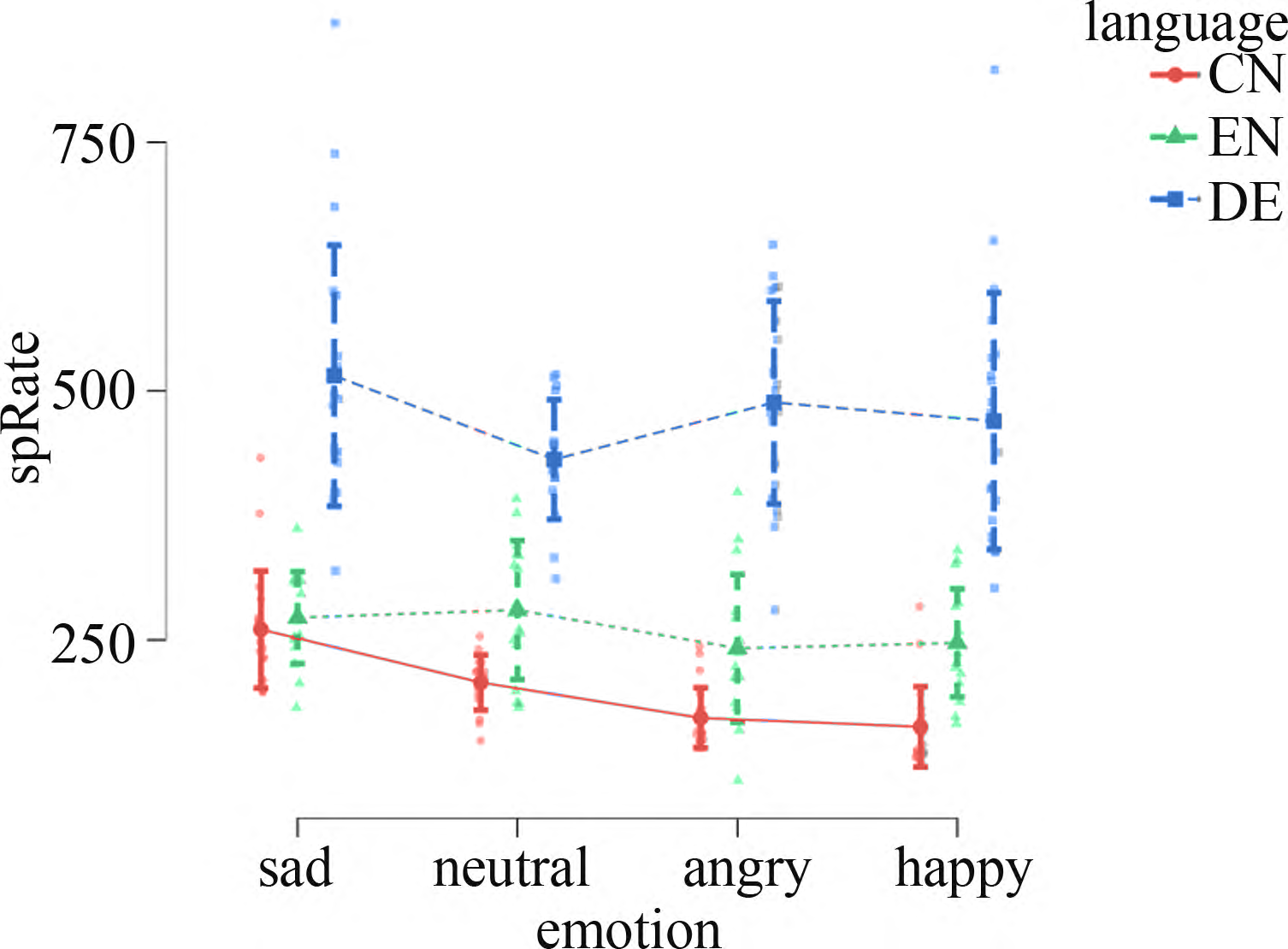}
\caption{spRate in Chinese, German, and English by emotion.}
\label{fig:5}
\end{figure}

As shown in Figure~\ref{fig:5}, in the Chinese dataset, the average spRate, from slowest to fastest, follows the order of sad, neutral, angry, and happy. In the English dataset, the slowest average spRate is observed in neutral, followed by sad, with those under happy and angry being faster in a similar manner. In the German dataset, the slowest spRate occurs in sad, followed by angry and happy, with the fastest appearing in neutral.

\begin{table}[!htbp]
\centering
\caption{ANOVA Results for the Effect of Emotion on spRate Across Languages.}
\label{tab:5}
\small
\renewcommand{\arraystretch}{1.12}
\setlength{\tabcolsep}{10pt}
\begin{tabular}{lrrr}
\toprule
language & sumsq & statistic & p.value \\
\midrule
CN & 117841 & 22.6 & \ensuremath{<}0.001 \\
DE & 66438 & 1.79 & 0.158 \\
EN & 17014 & 1.47 & 0.232 \\
\bottomrule
\end{tabular}
\end{table}

The ANOVA results (Table~\ref{tab:5}) revealed a significant effect of emotion on spRate in the Chinese dataset, indicating that emotional state plays a critical role in influencing spRate in Chinese. However, no significant effects of emotion on spRate were observed in the German or English datasets. These findings suggest that emotional variation has a notable impact on speech rate in Chinese, while its influence appears negligible in German and English.

A mixed-effects model was conducted to analyze the effect of emotion on spRate. The intercept was significant (\ensuremath{\beta} = 307.69, SE = 86.86, t = 3.54, p \ensuremath{<} 0.001). Among the emotional predictors, only sad showed a significant positive effect on speech rate (\ensuremath{\beta} = 45.00, SE = 15.66, t = 2.87, p = 0.004), suggesting that the speech rate increased under the sad emotion. The other emotional predictors, angry (\ensuremath{\beta} = -5.13, SE = 15.66, t = -0.33, p = 0.743) and happy (\ensuremath{\beta} = -13.14, SE = 15.67, t = -0.84, p = 0.402), did not significantly influence speech rate. For random effects, the variance components were substantial for language ($\tau_{00}$ = 22,105.07) and relatively small for gender ($\tau_{00}$ = 84.33), with a residual variance of 6332.15. The ICC is 0.78, indicating that a large portion of the variability in spRate is attributed to differences across languages and gender. The marginal $R^2$ is 0.018, showing a limited explanatory power of fixed effects, while the conditional $R^2$ is 0.782, reflecting the strong contribution of random effects to the model. In a word, while language and gender significantly account for variability in spRate, the influence of emotional state is limited, with only sad emotion showing a notable effect.

\subsection{Speech emotion recognition}

\subsubsection{Model performance}

\begin{figure}[!htbp]
\centering
\includegraphics[width=.64\linewidth]{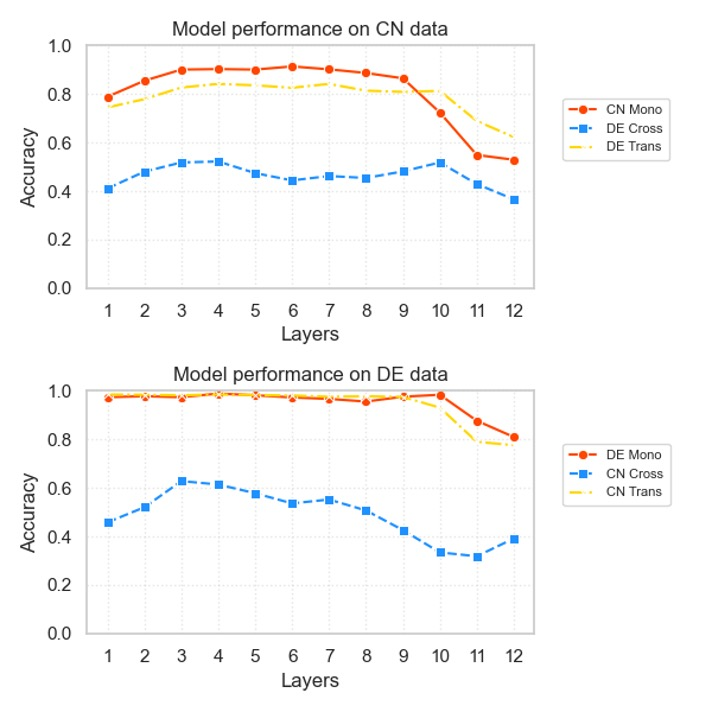}
\caption{Layer-wise analysis of CN and DE models under monolingual, cross-lingual and transfer learning settings.}
\label{fig:6}
\end{figure}

Figure~\ref{fig:6} illustrates the results of the layer-wise analysis. In the monolingual setting, both the CN and DE models perform strongly in their respective source languages, which aligns with expectations given their pre-training in these languages. However, in the cross-lingual setting, a substantial decline in accuracy is observed for both models. Although some reduction is expected due to language differences, the magnitude of this decline exceeds initial expectations, particularly considering the shared properties of emotional acoustics (Jeon et al., 2013).

A plausible explanation is that SSL models not only capture low-level acoustic features but also transform them into high-level linguistic information, such as word identity and meaning (Pasad et al., 2021). This transformation introduces a linguistic gap across languages, which amplifies the decline in accuracy. However, in a transfer learning setting, the models can achieve performance levels comparable to the monolingual scenario, highlighting the adaptability of SSL models for SER tasks across languages when effective knowledge transfer techniques are applied. The observed variations in contours are likely influenced by the training objectives of SSL models, particularly contrastive masked segment prediction.

\subsubsection{Model vs. Human}

To investigate whether SSL-based models can achieve SER performance on par with human evaluators, we conducted a comparative analysis by contrasting the unweighted accuracy of models and human subjects.

\paragraph{1) Monolingual}

As shown in Table~\ref{tab:6} and Table~\ref{tab:7}, both models outperform their respective native human subjects in terms of overall UA. A closer examination of the confusion matrices for the CN and DE monolingual models, as well as for CN and DE native speakers across all emotion categories as shown in Table~\ref{tab:8}, reveals that DE natives are more likely to misclassify neutral DE speech as sad than the DE monolingual model. In contrast, CN natives show lower precision in identifying happy and neutral emotions than the CN monolingual model. These findings suggest that SSL models can achieve strong monolingual performance on the SER task when provided with adequate training data.

\begin{table}[!htbp]
\centering
\caption{Model performance under monolingual, cross-lingual, and transfer learning with various PEFT strategies.}
\label{tab:6}
\small
\renewcommand{\arraystretch}{1.12}
\setlength{\tabcolsep}{8pt}
\begin{tabular}{llllcccr}
\toprule
Model & Setting & Source & Target & \multicolumn{3}{c}{PEFT strategy} & UA\% \\
\cmidrule(lr){5-7}
 & & & & LoRa & BA+WG & 2-stg & \\
\midrule
W2V2-CN & Mono & CN & CN &  &  &  & 91.4 \\
 &  &  &  & $\checkmark$ &  &  & 87.0 \\
 &  &  &  & $\checkmark$ & $\checkmark$ &  & 93.9 \\
\midrule
 & Cross & CN & DE &  &  &  & 62.8 \\
 &  &  &  & $\checkmark$ &  &  & 65.3 \\
 &  &  &  & $\checkmark$ & $\checkmark$ &  & 70.7 \\
\midrule
 & Trans & DE & DE &  &  &  & 98.5 \\
 &  &  &  & $\checkmark$ &  &  & 98.8 \\
 &  &  &  & $\checkmark$ & $\checkmark$ &  & 98.8 \\
 &  &  &  & $\checkmark$ & $\checkmark$ & $\checkmark$ & 98.9 \\
\midrule
 & Trans & EN & EN &  &  &  & 65.5 \\
 &  &  &  & $\checkmark$ &  &  & 66.3 \\
 &  &  &  & $\checkmark$ & $\checkmark$ &  & 67.7 \\
\midrule
W2V2-DE & Mono & DE & DE &  &  &  & 98.9 \\
 &  &  &  & $\checkmark$ &  &  & 97.8 \\
 &  &  &  & $\checkmark$ & $\checkmark$ &  & 97.9 \\
\midrule
 & Cross & DE & CN &  &  &  & 52.2 \\
 &  &  &  & $\checkmark$ &  &  & 58.5 \\
 &  &  &  & $\checkmark$ & $\checkmark$ &  & 56.0 \\
\midrule
 & Trans & CN & CN &  &  &  & 84.2 \\
 &  &  &  & $\checkmark$ &  &  & 83.6 \\
 &  &  &  & $\checkmark$ & $\checkmark$ &  & 87.5 \\
 &  &  &  & $\checkmark$ & $\checkmark$ & $\checkmark$ & 85.8 \\
\midrule
 & Trans & EN & EN &  &  &  & 62.4 \\
 &  &  &  & $\checkmark$ &  &  & 65.0 \\
 &  &  &  & $\checkmark$ & $\checkmark$ &  & 66.0 \\
\bottomrule
\end{tabular}
\end{table}

\paragraph{2) Cross-lingual}

In terms of overall accuracy, as shown in Tables 6 and 7, both humans and models experience a decrease in performance in the cross-lingual condition, with cross-lingual models being more significantly impacted than humans, which indicates that humans handle cross-lingual scenarios more effectively. Regarding performance across all emotion categories, as illustrated in Table~\ref{tab:8}, the DE cross-lingual model has difficulty in recognizing neutral and sad in CN data, leading to low recall rates. Furthermore, this model tends to confuse anger and happy more than human participants do in both languages. On the other hand, the CN cross-lingual model demonstrates a strong alignment with CN native speakers when interpreting DE speech, with both groups frequently misclassifying happy as neutral.

\begin{table}[!htbp]
\centering
\caption{Human performance (UA\%) in all languages (Human participants listened to the same speech samples as the model and classified the emotions).}
\label{tab:7}
\small
\renewcommand{\arraystretch}{1.12}
\setlength{\tabcolsep}{10pt}
\begin{tabular}{lrrr}
\toprule
 & CN & DE & EN \\
\midrule
CN participants & 79.5 & 73.3 & 63.5 \\
DE participants & 82.6 & 91.7 & 73.6 \\
\bottomrule
\end{tabular}
\end{table}

\begin{table}[!htbp]
\centering
\caption{Confusion matrices of CN and DE speakers and models under monolingual (first column), cross-lingual (second column), and transfer learning (third column) settings. No humans under the transfer learning setting (A-angry; H-happy; N-neutral; S-sad).}
\label{tab:8}
\small
\renewcommand{\arraystretch}{1.12}
\setlength{\tabcolsep}{3.5pt}
\begin{minipage}[t]{.325\linewidth}\centering
\begin{tabular}{lrrrr}
\toprule
 & A & H & N & S \\
A & 0.71 & 0.24 & 0.06 & 0.00 \\
H & 0.00 & 0.82 & 0.18 & 0.00 \\
N & 0.00 & 0.10 & 0.86 & 0.04 \\
S & 0.00 & 0.01 & 0.18 & 0.81 \\
\bottomrule
\end{tabular}
\par\smallskip {\scriptsize (a) Human: CN (mono)}
\end{minipage}\hfill
\begin{minipage}[t]{.325\linewidth}\centering
\begin{tabular}{lrrrr}
\toprule
 & A & H & N & S \\
A & 0.81 & 0.13 & 0.06 & 0.01 \\
H & 0.21 & 0.38 & 0.42 & 0.00 \\
N & 0.01 & 0.01 & 0.93 & 0.04 \\
S & 0.01 & 0.00 & 0.15 & 0.83 \\
\bottomrule
\end{tabular}
\par\smallskip {\scriptsize (e) Human: CN cross (on DE)}
\end{minipage}\hfill
\begin{minipage}[t]{.325\linewidth}\centering
\begin{tabular}{lrrrr}
\toprule
 & A & H & N & S \\
A & 0.54 & 0.22 & 0.13 & 0.11 \\
H & 0.14 & 0.61 & 0.19 & 0.06 \\
N & 0.04 & 0.28 & 0.58 & 0.10 \\
S & 0.03 & 0.01 & 0.15 & 0.81 \\
\bottomrule
\end{tabular}
\par\smallskip {\scriptsize (k) Human: CN (L2 on EN)}
\end{minipage}
\par\medskip
\begin{minipage}[t]{.325\linewidth}\centering
\begin{tabular}{lrrrr}
\toprule
 & A & H & N & S \\
A & 0.95 & 0.05 & 0.00 & 0.00 \\
H & 0.02 & 0.97 & 0.02 & 0.00 \\
N & 0.00 & 0.00 & 1.00 & 0.00 \\
S & 0.00 & 0.00 & 0.03 & 0.97 \\
\bottomrule
\end{tabular}
\par\smallskip {\scriptsize (b) Model: CN mono (on CN)}
\end{minipage}\hfill
\begin{minipage}[t]{.325\linewidth}\centering
\begin{tabular}{lrrrr}
\toprule
 & A & H & N & S \\
A & 0.95 & 0.02 & 0.03 & 0.00 \\
H & 0.08 & 0.18 & 0.51 & 0.22 \\
N & 0.00 & 0.00 & 0.95 & 0.05 \\
S & 0.00 & 0.00 & 0.00 & 1.00 \\
\bottomrule
\end{tabular}
\par\smallskip {\scriptsize (f) Model: CN cross (on DE)}
\end{minipage}\hfill
\begin{minipage}[t]{.325\linewidth}\centering
\begin{tabular}{lrrrr}
\toprule
 & A & H & N & S \\
A & 0.70 & 0.13 & 0.10 & 0.07 \\
H & 0.08 & 0.57 & 0.35 & 0.00 \\
N & 0.07 & 0.18 & 0.75 & 0.00 \\
S & 0.00 & 0.05 & 0.02 & 0.93 \\
\bottomrule
\end{tabular}
\par\smallskip {\scriptsize (l) Model: CN trans (L2 on EN)}
\end{minipage}
\par\medskip
\begin{minipage}[t]{.325\linewidth}\centering
\begin{tabular}{lrrrr}
\toprule
 & A & H & N & S \\
A & 0.97 & 0.03 & 0.00 & 0.00 \\
H & 0.00 & 0.94 & 0.05 & 0.00 \\
N & 0.00 & 0.00 & 0.83 & 0.17 \\
S & 0.00 & 0.00 & 0.04 & 0.96 \\
\bottomrule
\end{tabular}
\par\smallskip {\scriptsize (c) Human: DE (mono)}
\end{minipage}\hfill
\begin{minipage}[t]{.325\linewidth}\centering
\begin{tabular}{lrrrr}
\toprule
 & A & H & N & S \\
A & 0.88 & 0.07 & 0.06 & 0.00 \\
H & 0.08 & 0.75 & 0.15 & 0.01 \\
N & 0.03 & 0.03 & 0.76 & 0.18 \\
S & 0.00 & 0.03 & 0.06 & 0.92 \\
\bottomrule
\end{tabular}
\par\smallskip {\scriptsize (g) Human: DE cross (on CN)}
\end{minipage}\hfill
\begin{minipage}[t]{.325\linewidth}\centering
\begin{tabular}{lrrrr}
\toprule
 & A & H & N & S \\
A & 0.78 & 0.03 & 0.04 & 0.15 \\
H & 0.06 & 0.69 & 0.22 & 0.03 \\
N & 0.15 & 0.13 & 0.60 & 0.13 \\
S & 0.01 & 0.01 & 0.10 & 0.88 \\
\bottomrule
\end{tabular}
\par\smallskip {\scriptsize (m) Human: DE (L2 on EN)}
\end{minipage}
\par\medskip
\begin{minipage}[t]{.325\linewidth}\centering
\begin{tabular}{lrrrr}
\toprule
 & A & H & N & S \\
A & 0.98 & 0.02 & 0.00 & 0.00 \\
H & 0.02 & 0.98 & 0.00 & 0.00 \\
N & 0.00 & 0.00 & 0.93 & 0.07 \\
S & 0.00 & 0.00 & 0.00 & 1.00 \\
\bottomrule
\end{tabular}
\par\smallskip {\scriptsize (d) Model: DE mono (on DE)}
\end{minipage}\hfill
\begin{minipage}[t]{.325\linewidth}\centering
\begin{tabular}{lrrrr}
\toprule
 & A & H & N & S \\
A & 0.52 & 0.48 & 0.00 & 0.00 \\
H & 0.17 & 0.65 & 0.17 & 0.02 \\
N & 0.00 & 0.38 & 0.38 & 0.23 \\
S & 0.05 & 0.28 & 0.30 & 0.37 \\
\bottomrule
\end{tabular}
\par\smallskip {\scriptsize (h) Model: DE cross (on CN)}
\end{minipage}\hfill
\begin{minipage}[t]{.325\linewidth}\centering
\begin{tabular}{lrrrr}
\toprule
 & A & H & N & S \\
A & 0.70 & 0.12 & 0.17 & 0.02 \\
H & 0.00 & 0.65 & 0.33 & 0.02 \\
N & 0.08 & 0.30 & 0.58 & 0.03 \\
S & 0.00 & 0.00 & 0.25 & 0.75 \\
\bottomrule
\end{tabular}
\par\smallskip {\scriptsize (n) Model: DE trans (on EN)}
\end{minipage}
\end{table}

Additionally, a two-sided Welch's t-test on the precision, recall, and F1 scores of humans was conducted. The result demonstrated significant differences in the recall of happy on the DE data between CN and DE subjects [ t (10) = -7.511, p \ensuremath{<} 0.001 ], as well as in the precision of neutral [ t (10) = -5.614, p \ensuremath{<} 0.001 ]. Furthermore, CN subjects showed lower recall for happy in DE data compared to CN data [ t (10) = -5.137, p \ensuremath{<} 0.001], suggesting a linguistic and paralinguistic knowledge gap between the two speaker groups. Particularly, significant differences were found in the recall of sad across CN, DE, and EN data [ t (10) = -2.708, p = 0.022] and in the precision of neutral [ t (10) = -7.511, p \ensuremath{<} 0.001 ]. The precision of neutral was notably affected by CN subjects' difficulty in perceiving happy in DE data, highlighting the impact of linguistic and paralinguistic differences on the perception of sad across languages.

\paragraph{3) Transfer learning}

The transfer learning setting, which mirrors the human process of acquiring a second language, allows for a comparison between SSL models and human subjects using EN data. As shown in Table~\ref{tab:6}, SSL models with transfer learning achieve monolingual-level performance and outperform human subjects on CN and DE data. On the EN data, DE subjects show higher accuracy than both CN subjects and all tested models. Additionally, two-stage fine-tuning does not significantly improve performance. These results suggest that while transfer learning assists SSL models in adapting to new languages, performance varies based on the specific characteristics of the target language dataset.

When analyzing performance across emotion categories, CN subjects only surpass the model in recognizing happy, whereas the CN transfer learning model outperforms humans in the other three emotion categories. For DE subjects, humans perform better in predicting happy and neutral emotions compared to the DE transfer learning model. Additionally, a PEFT strategy that proves effective in monolingual scenarios does not necessarily yield benefits in cross-lingual or multilingual contexts. Table~\ref{tab:7} further indicates that emotion recognition in EN is more challenging than in CN and DE despite CN and DE subjects being L2 learners. This increased difficulty can likely be attributed to the use of only improvised utterances from the IEMOCAP dataset, which represent more natural, real-life emotions, making the SER task more challenging.

\subsection{Speech emotion diarization}

To investigate whether models can identify emotionally salient segments in a manner analogous to human behavior, we compared model performance with that of human subjects. Furthermore, to elucidate how human listeners leverage prosodic features to recognize emotions, particularly when processing non-native speech, we conducted a detailed analysis of the acoustic characteristics of the segments selected by the participants.

\subsubsection{Model vs. Human}

The results presented in Table~\ref{tab:9} reveal that in general both human groups outperform the model, with the DE group achieving the lowest Emotion Diarization Error Rate (EDER). The model performs best in recognizing happiness and worst in identifying sad. Among the human groups, CN subjects are slightly more accurate in perceiving anger, while DE subjects perform better in identifying sad. This pattern aligns with the Speech Emotion Recognition results as shown in Table~\ref{tab:8}, where CN subjects exhibit a higher threshold for recognizing sad, leading to higher recall but lower precision. In contrast, DE subjects show higher precision but lower recall. The difference in sensitivity to sad among CN subjects results in a greater number of false negatives for sad in the SED task.

\begin{table}[!htbp]
\centering
\caption{EDER (\%) comparison of WavLM and humans on ZED data.}
\label{tab:9}
\small
\renewcommand{\arraystretch}{1.12}
\setlength{\tabcolsep}{10pt}
\begin{tabular}{lrrr}
\toprule
 & WavLM & CN participants & DE participants \\
\midrule
Angry & 36.6 & 25.8 & 27.5 \\
Happy & 27.5 & 31.8 & 28.7 \\
Sad & 50.3 & 38.6 & 28.3 \\
Average & 38.2 & 32.1 & 28.2 \\
\bottomrule
\end{tabular}
\end{table}

\subsubsection{Acoustic features of the selected segments}

After completing the SED task, we conducted a more detailed examination of the speech segments selected by human participants. The primary acoustic variables investigated included the intMax, intMin, intMean, $f_0$ Max, $f_0$ Min, and $f_0$ Mean, as well as the duration of the selected segments. The pitch-related variables were normalized using a semitone scale, while the intensity-related variables were normalized using the min-max method. The duration of the selected segments was calculated as the proportion of the segment's length relative to the entire sentence.

\paragraph{1) Duration of selected segments}

\begin{figure}[!htbp]
\centering
\includegraphics[width=.70\linewidth]{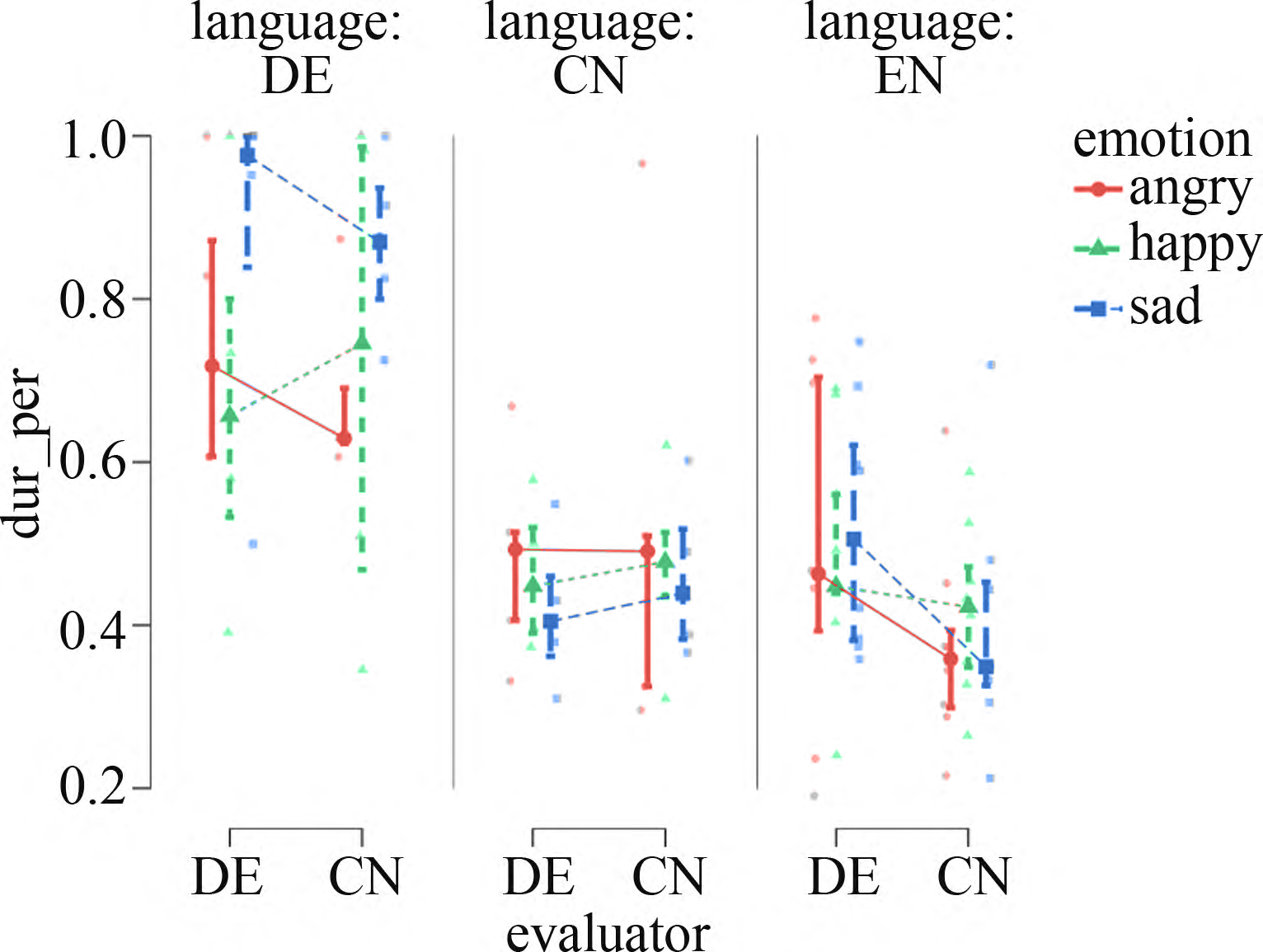}
\caption{Duration of selected segments by subjects and emotion under various datasets.}
\label{fig:7}
\end{figure}

In Figure~\ref{fig:7}, dur\_per represents the proportion of the selected segment's duration relative to the entire sentence's duration.

Across both Chinese and German listeners, German stimuli consistently required greater processing difficulty than Chinese or English stimuli, as indicated by the larger proportion of selected-segment durations.

German participants' response time patterns when perceiving English, a language typologically similar to German itself, were similar to those observed when listening to their native language, displaying the ordering sad \ensuremath{>} angry \ensuremath{>} happy. However, their overall response times were shorter for English than for German stimuli. The difference may stem from the pronunciation characteristics of the German speakers who produced the recordings. In contrast, when listening to Chinese, a tonal language that differs significantly from German, they exhibited a markedly different response-time pattern.

For individual emotions, in the case of German stimuli, Chinese participants spent the most time on segments marked as sad, followed by happy, with the least time spent on angry. Similarly, German participants also spent the most time on sad, followed by angry, and the least on happy.

For Chinese stimuli, both Chinese and German participants displayed similar temporal patterns: the most time was spent on angry, followed by happy, and the least on sad. However, Chinese participants tended to select relatively longer speech segments when processing Chinese stimuli.

For English stimuli, German participants spent the most time on sad, followed by angry, and the least on happy. In contrast, Chinese participants spent the most time on happy, followed by angry, and the least on sad. Notably, when processing English stimuli, Chinese participants generally selected shorter speech segments compared to other languages.

An ANOVA test was conducted with language, emotion, and evaluator as fixed factors. The analysis revealed a significant main effect of language [F (2, 81) = 28.05, p \ensuremath{<} 0.001]. However, neither emotion nor evaluator showed significant main effects (both p \ensuremath{>} 0.40). Although the descriptive plot suggests numerical differences across evaluators for some emotion categories, these differences were not statistically reliable.

\paragraph{2) Chinese}

\begin{figure}[!htbp]
\centering
\includegraphics[width=1.0\linewidth]{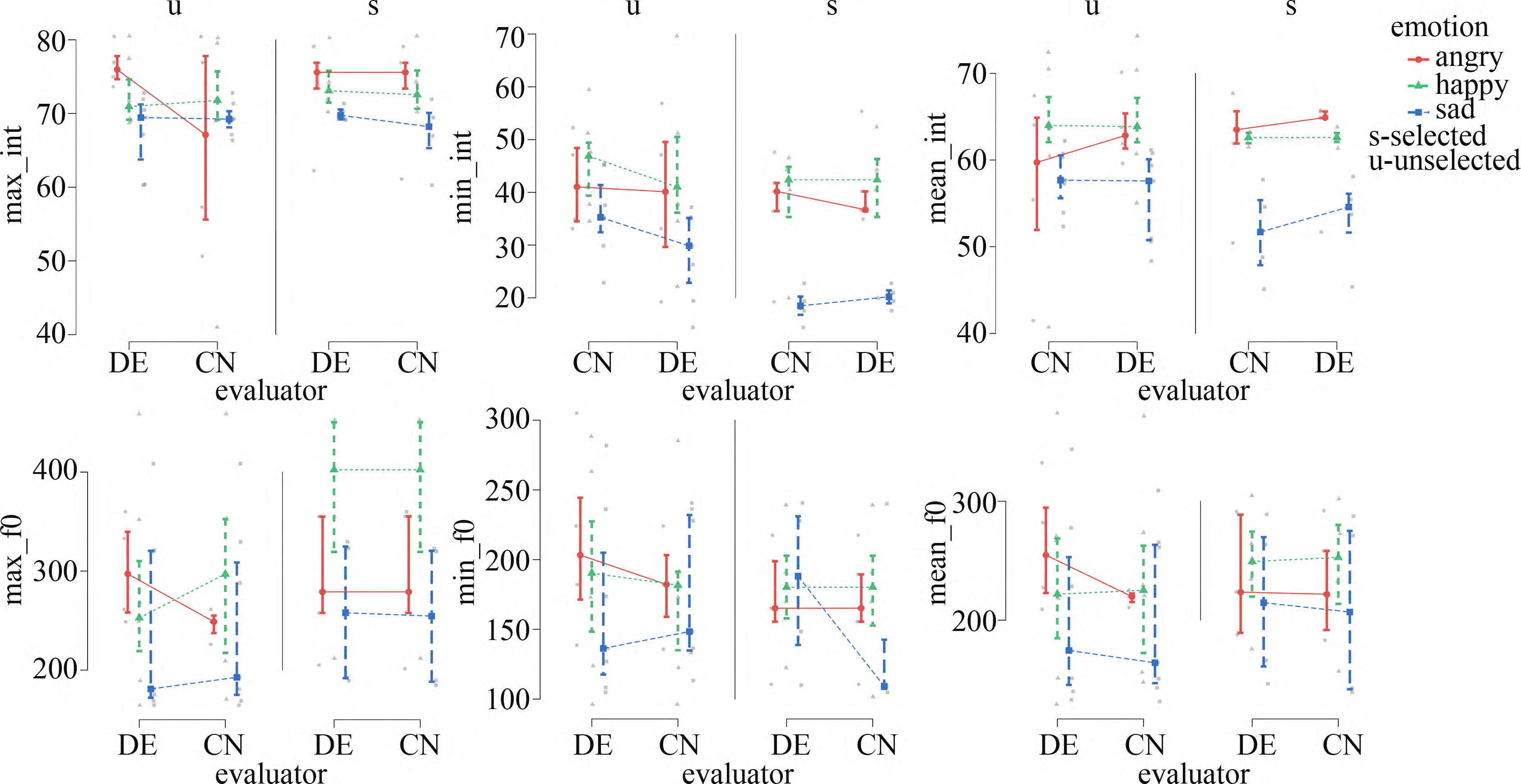}
\caption{Comparison of acoustic parameters by subjects, selection status, and emotion under Chinese.}
\label{fig:8}
\end{figure}

When perceiving emotions in Chinese, the speech segments selected by participants from both countries exhibited similar patterns in intensity-related features, as shown in Figure~\ref{fig:8}: in both max\_int and mean\_int group, the strongest intensity was observed for angry, followed by happy, and the weakest for sad. While in unselected segments, minimum and mean intensity also exhibited similar patterns for Chinese and German participants, suggesting a comparable role of these parameters in their emotion perception.

In terms of fundamental frequency, while the maximum $f_0$ of selected segments varies noticeably across emotions, the average maximum $f_0$ for happy in segments chosen by both groups is significantly higher than that of unselected segments. Regarding minimum $f_0$, Chinese participants appear more likely to interpret lower values as indicative of sad.

ANOVA test results, as shown in Table~\ref{tab:10}, revealed that emotion exerted significant effects on several acoustic measures. For intensity-related features, emotion significantly influenced min\_int and mean\_int, whereas its effect on max\_int did not reach significance. For $f_0$-related parameters, emotion yielded a significant effect on max\_f0, but not on min\_f0 or mean\_f0.

In contrast, evaluator and selection status showed no significant effects on any of the acoustic measures, except for a modest effect of selection status on min\_int. Overall, the results indicate that emotion serves as the primary factor shaping acoustic variation, whereas the evaluators' language background or selection status differences contribute minimally.

\begin{table}[!htbp]
\centering
\caption{ANOVA results for the effects of emotion, evaluator, and selection status on the acoustic parameters under Chinese.}
\label{tab:10}
\small
\renewcommand{\arraystretch}{1.12}
\setlength{\tabcolsep}{5pt}
\begin{tabular}{llrrrrr}
\toprule
 & Cases & Sum of Squares & df & Mean Square & F & p \\
\midrule
max\_int & emotion & 200.71 & 2 & 100.36 & 1.929 & 0.156 \\
 & evaluator & 93.44 & 1 & 93.44 & 1.796 & 0.186 \\
 & selection status & 47.23 & 1 & 47.23 & 0.908 & 0.345 \\
\midrule
min\_int & emotion & 3018.48 & 2 & 1509.24 & 13.691 & \ensuremath{<} .001 \\
 & evaluator & 11.85 & 1 & 11.85 & 0.108 & 0.744 \\
 & selection status & 669.36 & 1 & 669.36 & 6.072 & 0.017 \\
\midrule
mean\_int & emotion & 810.924 & 2 & 405.462 & 9.542 & \ensuremath{<} .001 \\
 & evaluator & 34.317 & 1 & 34.317 & 0.808 & 0.373 \\
 & selection status & 19.158 & 1 & 19.158 & 0.451 & 0.505 \\
\midrule
max\_f0 & emotion & 56890.2 & 2 & 28445.1 & 3.49 & 0.038 \\
 & evaluator & 602.7 & 1 & 602.7 & 0.074 & 0.787 \\
 & selection status & 17514.8 & 1 & 17514.8 & 2.149 & 0.149 \\
\midrule
min\_f0 & emotion & 2934.48 & 2 & 1467.24 & 0.413 & 0.664 \\
 & evaluator & 2683.12 & 1 & 2683.12 & 0.755 & 0.389 \\
 & selection status & 2949.28 & 1 & 2949.28 & 0.829 & 0.367 \\
\midrule
mean\_f0 & emotion & 10400.7 & 2 & 5200.4 & 1.098 & 0.342 \\
 & evaluator & 1583.2 & 1 & 1583.2 & 0.334 & 0.566 \\
 & selection status & 149.3 & 1 & 149.3 & 0.032 & 0.86 \\
\bottomrule
\end{tabular}
\end{table}

\paragraph{3) German}

\begin{figure}[!htbp]
\centering
\includegraphics[width=1.0\linewidth]{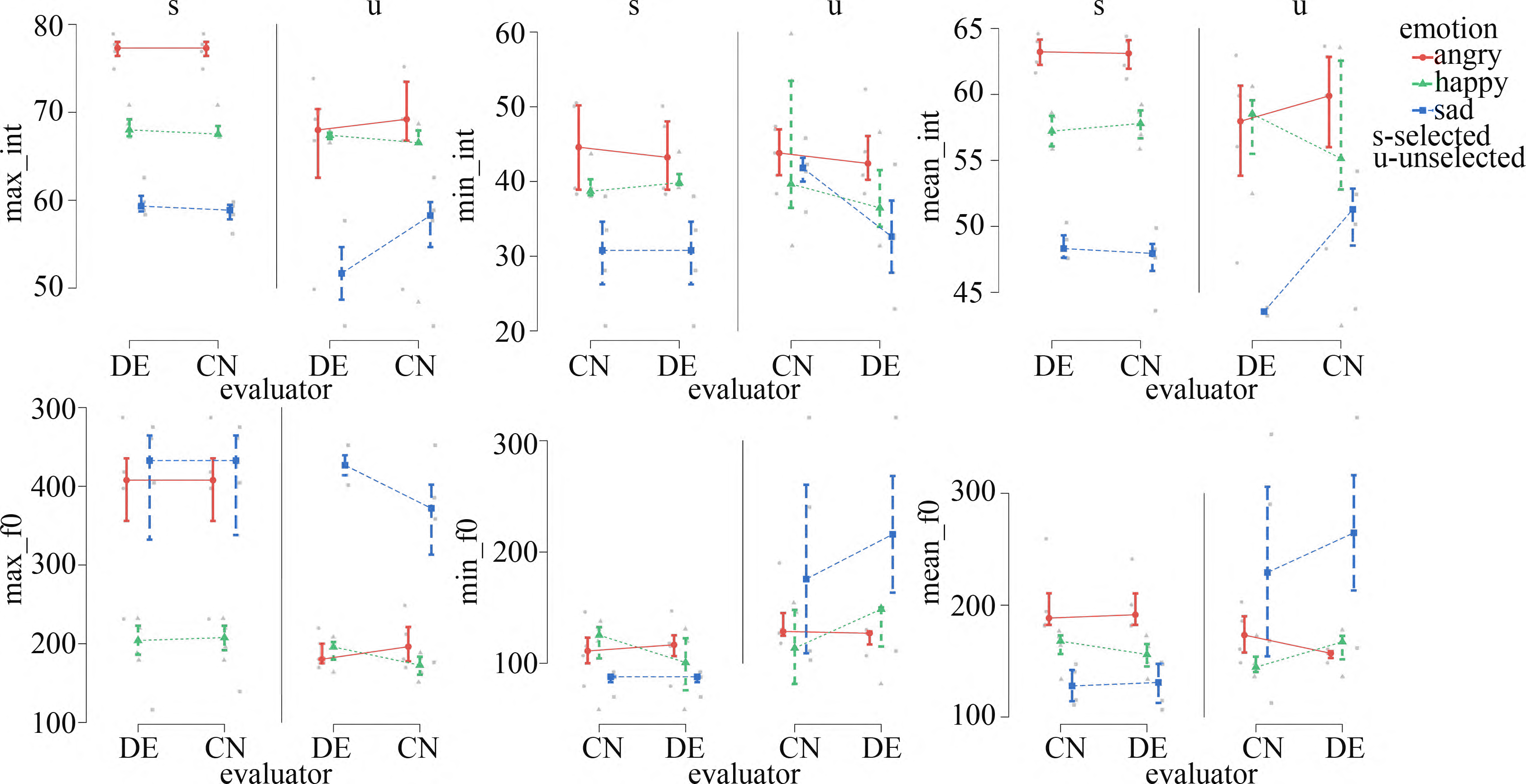}
\caption{Comparison of acoustic parameters by subjects, selection status, and emotion under German.}
\label{fig:9}
\end{figure}

\begin{table}[!htbp]
\centering
\caption{ANOVA results for the effects of emotion, evaluator, and selection status on the acoustic parameters under German.}
\label{tab:11}
\small
\renewcommand{\arraystretch}{1.12}
\setlength{\tabcolsep}{5pt}
\begin{tabular}{llrrrrr}
\toprule
 & Cases & Sum of Squares & df & Mean Square & F & p \\
\midrule
max\_int & emotion & 1664.305 & 2 & 832.152 & 22.164 & \ensuremath{<} .001 \\
 & evaluator & 0.38 & 1 & 0.38 & 0.01 & 0.92 \\
 & selection status & 463.957 & 1 & 463.957 & 12.357 & 0.001 \\
\midrule
min\_int & emotion & 799.55 & 2 & 399.78 & 8.178 & 0.001 \\
 & evaluator & 62.06 & 1 & 62.06 & 1.269 & 0.268 \\
 & selection status & 67.93 & 1 & 67.93 & 1.389 & 0.246 \\
\midrule
mean\_int & emotion & 1237.393 & 2 & 618.697 & 29.319 & \ensuremath{<} .001 \\
 & evaluator & 8.472 & 1 & 8.472 & 0.401 & 0.53 \\
 & selection status & 80.663 & 1 & 80.663 & 3.822 & 0.059 \\
\midrule
max\_f0 & emotion & 230800 & 2 & 115400 & 13.065 & \ensuremath{<} .001 \\
 & evaluator & 1935 & 1 & 1935 & 0.219 & 0.643 \\
 & selection status & 44421 & 1 & 44421 & 5.029 & 0.032 \\
\midrule
min\_f0 & emotion & 7097.535 & 2 & 3548.767 & 1.431 & 0.254 \\
 & evaluator & 1.929 & 1 & 1.929 & $7.779\times 10^{-4}$ & 0.978 \\
 & selection status & 28195.373 & 1 & 28195.373 & 11.369 & 0.002 \\
\midrule
mean\_f0 & emotion & 8500.2 & 2 & 4250.1 & 1.889 & 0.168 \\
 & evaluator & 78.5 & 1 & 78.5 & 0.035 & 0.853 \\
 & selection status & 7099.69 & 1 & 7099.69 & 3.155 & 0.085 \\
\bottomrule
\end{tabular}
\end{table}

As shown in Figure~\ref{fig:9}, for the German dataset, both Chinese and German participants selected speech segments that demonstrated a strong uniformity in terms of intensity. This suggests an agreement of the role of intensity in emotion recognition within German speech, with the strongest intensity observed under angry, followed by happy, and then sad. However, the differences in the unselected segments, particularly for sad emotions, indicate that native German subjects do not solely rely on intensity when perceiving emotional cues in their own language. In contrast, Chinese participants tend to select an entire segment with lower intensity, especially under sad emotions.

From the perspective of pitch-related features, the selected segments from both groups of participants also show similarities. Specifically, the average pitch is highest for angry, followed by happy, and then sad. It is noteworthy that the high dynamic range of pitch in German sad emotion (with a high $f_0$ Max and a low $f_0$ Min) may indicate a more complex manifestation of sad. This pattern encompasses both low-energy, elongated segments and occasional high-pitch emotional peaks. These characteristics in the selected segments suggest that, for both Chinese and German participants, the perception of sad in German may be associated with pitch range expansion.

From the ANOVA results shown in Table~\ref{tab:11}, robust effects of emotion on all three intensity measures were found. Emotion significantly influenced max\_int, min\_int, and mean\_int. Selection status also showed significant effects on max\_int, whereas its effects on min\_int and mean\_int did not reach significance. No effects of the evaluator were observed for any intensity parameter.

For the $f_0$-related measures, emotion represented a strong effect on max\_f0, but not on min\_f0 or mean\_f0. Selection status significantly affected min\_f0, while also showing a smaller effect on max\_f0. As with the intensity measures, evaluators' language backgrounds had no significant impact on any $f_0$ parameter. Together, these findings indicate that emotion is the primary source of acoustic variation, while selection status represents selective influence on certain parameters, and evaluator backgrounds contribute minimally.

\paragraph{4) English}

When identifying emotions in English, the segments selected by both Chinese and German participants exhibited notable differences in maximum intensity across emotional categories. The maximum intensity is higher for emotions such as angry and happy, while it is significantly lower for sad. Regarding the minimum intensity, German subjects tended to select segments with lower intensity as representative of sad emotions, suggesting that they relied more on lower intensity to indicate sad compared to their Chinese counterparts.

From the perspective of the fundamental frequency, participants from both groups showed a consistent overall pattern: the selected segments had the highest $f_0$ for happy, followed by angry, and the lowest for sad. However, German participants displayed a stronger preference for selecting segments with either more pronounced high or distinctly low $f_0$ values as representative of emotional expressions

ANOVA results for the English dataset are summarized in Table~\ref{tab:12}. For intensity-related measures, emotion had a significant effect only on min\_int, whereas its effects on maximum and mean intensity were not significant. Selection status significantly influenced maximum intensity, but not minimum or mean intensity. No effects of evaluator were observed for any intensity parameter.

For the $f_0$-related parameters, robust emotional effects were found across multiple measures. Emotion significantly affected maximum $f_0$, minimum $f_0$, and mean $f_0$. Selection status also showed a significant impact on maximum $f_0$, and mean $f_0$, though it did not affect minimum $f_0$. As with intensity measures, evaluator effects were non-significant for all $f_0$ parameters.

\begin{figure}[!htbp]
\centering
\includegraphics[width=1.0\linewidth]{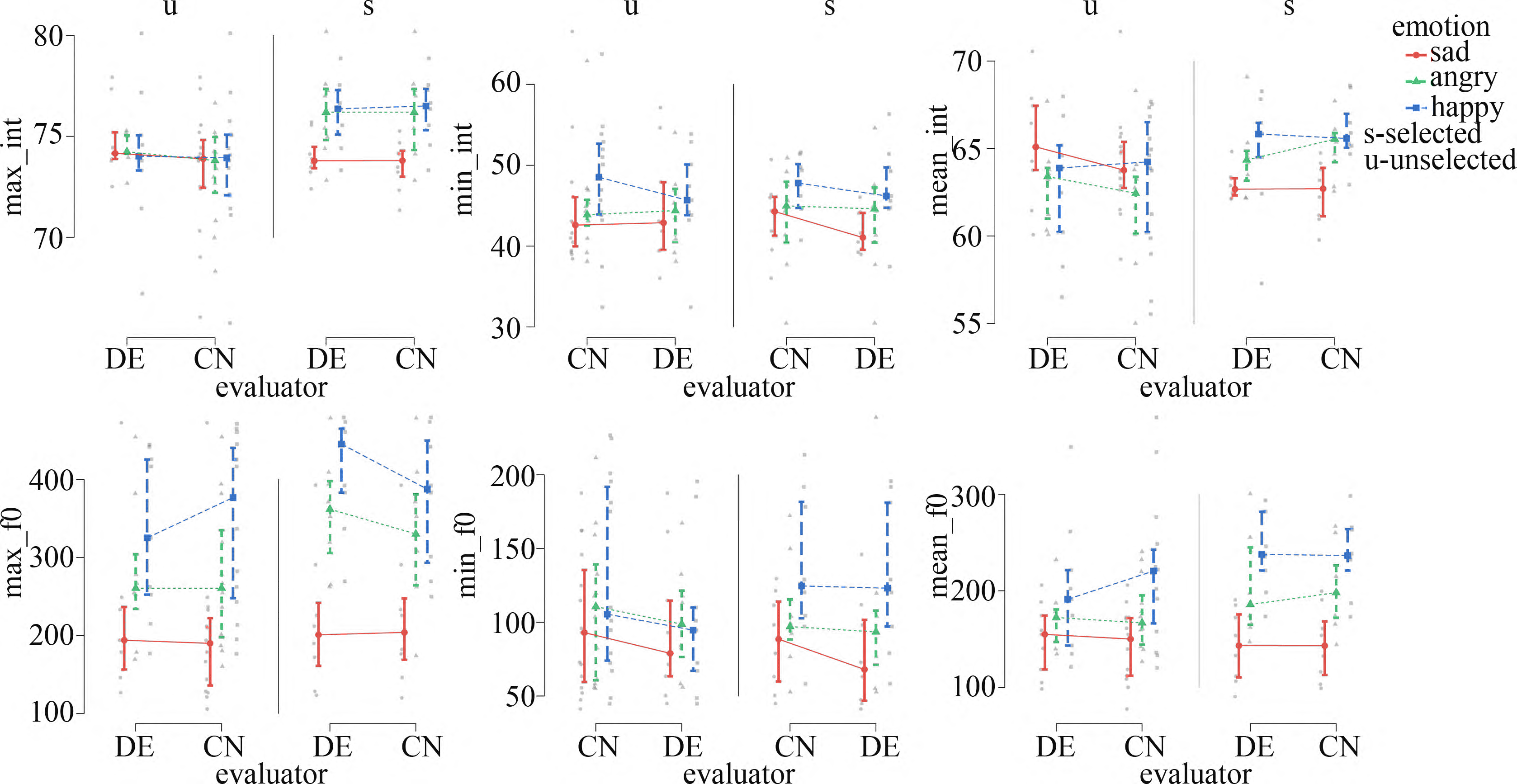}
\caption{Comparison of acoustic parameters by subjects, selection status, and emotion under English.}
\label{fig:10}
\end{figure}

\begin{table}[!htbp]
\centering
\caption{ANOVA results for the effects of emotion, evaluator, and selection status on the acoustic parameters under English.}
\label{tab:12}
\small
\renewcommand{\arraystretch}{1.12}
\setlength{\tabcolsep}{5pt}
\begin{tabular}{llrrrrr}
\toprule
 & Cases & Sum of Squares & df & Mean Square & F & p \\
\midrule
max\_int & emotion & 27.666 & 2 & 13.833 & 2.359 & 0.1 \\
 & evaluator & 8.848 & 1 & 8.848 & 1.509 & 0.222 \\
 & Selection status & 59.304 & 1 & 59.304 & 10.113 & 0.002 \\
\midrule
min\_int & emotion & 268.149 & 2 & 134.074 & 3.382 & 0.038 \\
 & evaluator & 13.582 & 1 & 13.582 & 0.343 & 0.56 \\
 & Selection status & 25.951 & 1 & 25.951 & 0.655 & 0.42 \\
\midrule
mean\_int & emotion & 8.22 & 2 & 4.11 & 0.473 & 0.624 \\
 & evaluator & 0.446 & 1 & 0.446 & 0.051 & 0.821 \\
 & Selection status & 24.067 & 1 & 24.067 & 2.772 & 0.099 \\
\midrule
max\_f0 & emotion & 438467.6 & 2 & 219234 & 24.723 & \ensuremath{<} .001 \\
 & evaluator & 4255.4 & 1 & 4255.41 & 0.48 & 0.49 \\
 & Selection status & 51632.7 & 1 & 51632.8 & 5.823 & 0.018 \\
\midrule
min\_f0 & emotion & 22787.251 & 2 & 11393.6 & 4.484 & 0.014 \\
 & evaluator & 2413.825 & 1 & 2413.83 & 0.95 & 0.332 \\
 & Selection status & 4.545 & 1 & 4.545 & 0.002 & 0.966 \\
\midrule
mean\_f0 & emotion & 119437.7 & 2 & 59718.8 & 25.134 & \ensuremath{<} .001 \\
 & evaluator & 117.4 & 1 & 117.4 & 0.049 & 0.825 \\
 & Selection status & 10132.8 & 1 & 10132.8 & 4.265 & 0.041 \\
\bottomrule
\end{tabular}
\end{table}

\section{Conclusion}

To explore the prosodic patterns of emotional expression across languages and to compare the performance of human and SSL models in cross-linguistic SER and SED tasks, this study first investigated the prosodic patterns the three languages (i.e., CN, DE, EN) employ to express emotions, with a particular focus on fundamental frequency and intensity. Using ANOVA and mixed-effects models, we examined the influence of emotions on these parameters while controlling for individual differences in gender and language. The findings revealed nuanced cross-linguistic and emotion-specific patterns, emphasizing the complex interplay between language, cultural factors, and emotional prosody.

After that, we examined the performance of cross-lingual SER between humans and SSL models. By comparing both monolingual and cross-lingual scenarios, as well as transferring learning settings, the results illustrated that humans consistently outperform SSL models in cross-lingual SER tasks. However, with the application of PEFT strategies and layer-wise analyses, SSL models showed significant improvement, achieving near-native performance in target languages. These findings highlight the potential of SSL models to adapt to cross-lingual SER through advanced fine-tuning techniques.

The study also explored SED, focusing on fine-grained recognition of emotionally salient speech segments. Prosodic analyses of the selected corpora revealed the distinct strategies employed by humans and SSL models in detecting segment-level emotions. While humans excelled in accurately identifying emotional segments across languages, SSL models showed the ability to adapt and capture segment-level emotions with remarkable precision, especially under transfer learning scenarios. These findings emphasize the complementary strengths of human perception and machine learning in understanding and processing emotional cues at the segment level.

Further analysis of the prosodic patterns in speech segments selected during the SED task yields the following conclusions: First, both Chinese and German participants demonstrated a shared reliance on intensity-related features across different languages, underscoring the universal significance of intensity as a cue for emotion perception (Juslin \& Laukka, 2001; Xiong et al., 2008). Meanwhile, notable differences observed in the unselected segments suggest that the acoustic parameters examined in this study alone are insufficient to fully capture the nuanced emotional cues perceived by native speakers, highlighting the potential role of additional acoustic or linguistic factors. Moreover, pitch-related features revealed cross-linguistic commonalities. For instance, in all languages, the $f_0$ was consistently the highest for happy, followed by angry, and the lowest for sad. However, the interpretation of pitch features varied across cultural and linguistic contexts. These findings suggest that emotion perception is influenced by both universal acoustic features and language-specific factors. Future research should incorporate more diverse datasets and control for variables such as speaker variation, cultural background, and language typology to refine these conclusions.

\FloatBarrier

\begingroup
\makeatletter
\renewcommand{\@biblabel}[1]{}
\makeatother

\endgroup

\FloatBarrier

\section*{Author information}
\small

\par\medskip\noindent\textbf{GENG Tianqi}\par

M.A. student at the Department of Languages, Cultures \& Applied Linguistics in Dietrich College of Humanities and Social Sciences at Carnegie Mellon University. Her research interests include speech prosody, laboratory phonology, and applied linguistics.

\noindent Email: \href{mailto:tianqig@andrew.cmu.edu}{tianqig@andrew.cmu.edu}

\par\medskip\noindent\textbf{HAN Zhichen}\par

M. S., graduated from the School of Philosophy, Psychology and Language Sciences, University of Edinburgh. His research interests include speech prosody and speech emotion recognition.

\noindent Email: \href{mailto:z.han@shisu.edu.cn}{z.han@shisu.edu.cn}

\par\medskip\noindent\textbf{FENG Hui}\par

Ph.D., Associate professor at the School of Foreign Languages, Tianjin University. Her research interests include speech prosody and laboratory phonology.

\noindent Email: \href{mailto:fenghui@tju.edu.cn}{fenghui@tju.edu.cn}

\par\medskip\noindent\textbf{YUAN Jiahong}\par

Ph.D., Professor at the School of Humanities and Social Sciences at the University of Science and Technology of China. His current research focuses on the early detection and prediction of Alzheimer's disease through spoken language, investigating the origin and evolution of Chinese tones with machine learning methods, and exploring model-brain alignment from the perspective of tone recognition.

\noindent Email: \href{mailto:jiahongyuan@ustc.edu.cn}{jiahongyuan@ustc.edu.cn}

\par\medskip\noindent\textbf{RICHMOND Korin}\par

Ph.D., Reader (Associate Professor) at the Centre for Speech Technology Research at the University of Edinburgh. His research interests include human language and speech technology, articulatory-acoustic relationship, and speech synthesis. He has served on the Organising Committee of multiple international and UK speech conferences.

\noindent Email: \href{mailto:korin@cstr.ed.ac.uk}{korin@cstr.ed.ac.uk}

\par\medskip\noindent\textbf{LI Yuanchao}\par

Ph.D., visiting researcher at the Interdisciplinary Research Center for Linguistic Sciences, University of Science and Technology of China. His research focuses on detecting mental, cognitive, and health states using speech-centric multimodal AI. His work has been recognized with awards from several associations including IEEE SPS, ACM SIGCHI, Alan Turing Institute, and Schmidt Sciences. He has (co-)organized multiple events at renowned speech conferences. He is the corresponding author of this paper.

\noindent Email: \href{mailto:yl809@cam.ac.uk}{yl809@cam.ac.uk}

\section*{Appendix}

The code used in this study is now available on:\\
\texttt{https://github.com/zhan7721/Crosslingual SER}

\end{document}